\documentclass[twocolumn,reprint,aps,prb,showkeys,groupaddress,superscriptaddress,sectionbib,square]{revtex4-1}
\setcitestyle{numbers,square}
\usepackage{epsfig}
\usepackage{color}
\begin{document}

\title{FeRh groundstate and martensitic transformation}

\author{Nikolai~A. Zarkevich}\email{zarkev@ameslab.gov}
\affiliation{Ames Laboratory, U.S. Department of Energy, Iowa State University, Ames, Iowa 50011-3020, USA}

\author{Duane~D. Johnson}\email{ddj@iastate.edu, ddj@ameslab.gov}
\affiliation{Ames Laboratory, U.S. Department of Energy, Iowa State University, Ames, Iowa 50011-3020, USA}
\affiliation{Department of Materials Science \& Engineering, Iowa State University, Ames, Iowa 50011, USA}

\date{\today}

\begin{abstract}
Cubic B2 FeRh exhibits a meta\-magnetic transition  [(111) anti\-ferro\-magnet (AFM) to ferro\-magnet (FM)] around 353~K and remains structurally stable at higher temperatures. However, the calculated zero-Kelvin phonons of AFM FeRh exhibit imaginary modes at M-points in the Brillouin zone, indicating a pre\-martensitic instability, which is a precursor to a  martensitic transformation at low temperatures.
Combining electronic-structure calculations with \textit{ab initio} molecular dynamics, conjugate gradient relaxation, and the solid-state nudged-elastic band (SSNEB) methods, we predict that AFM B2 FeRh becomes unstable at ambient pressure and transforms without a barrier to an AFM(111) orthorhombic (martensitic) groundstate below $90\pm10\,$K.
We also consider competing structures, in particular, a  tetragonal AFM(100) phase that is not the global ground\-state, as proposed  [Phys. Rev. B {\bf 94}, 180407(R) (2016)], but a constrained solution.
\end{abstract}

\maketitle


Discovery of a symmetry-breaking martensitic phase transition in such a well-studied magnetic inter\-metallic compound as FeRh presents a great scientific interest, while a giant magneto\-caloric effect at the metamagnetic transition near room temperature $T$ has a potential use in the solid-state refrigerators and heat pumps.

{\par}  FeRh cubic B2 phase with AFM(111) spin order is found to be unstable at ambient pressure  \cite{2017.FeRh.calorics,PRB94p014109y2016,PRB94p174435y2016,2016.PRB.94.180407}.
However, the stable structure and martensitic transformation path are unknown.
To establish these, we perform simulated annealing via \emph{ab initio} molecular dynamics,  and use the solid-state nudged-elastic band (SSNEB) methods \cite{SSNEB,JChemPhys142p064707y2015} within density-functional theory (DFT) to determine the minimum-enthalpy path (MEP) and enthalpy barrier, along with structural properties.

\begin{figure}[b!]
\includegraphics[width=80mm]{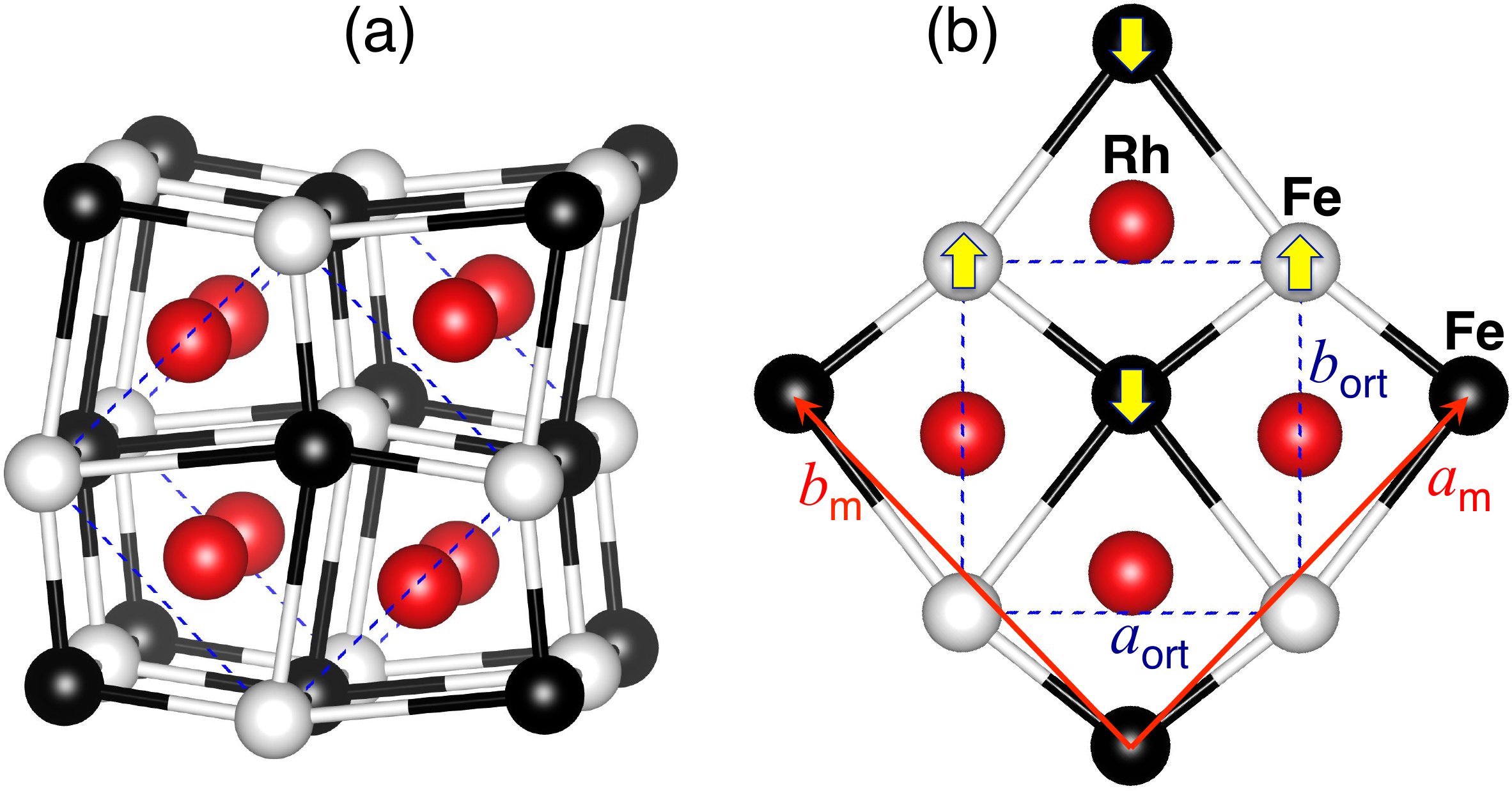} 
\caption{\label{fig1} (Color online). AFM FeRh orthorhombic ($Pmmn$) groundstate (a) and its (001) projection (b). Fe moments are oriented up (white) and down (black); Rh (red) moments are zero.
Lattice vectors of the primitive cell (dashed blue lines) are oriented along cubic [110], $[\bar{1}10]$, and [002] directions.
}
\end{figure}

{\par} The groundstate is an orthorhombic ($Pmmn$) structure with the type-2 AFM(111) spin order, as in Fig.~\ref{fig1} and Table~\ref{t1ort}.
 Magnetic moments of Rh are zero in a type-2 AFM structure (Fig.~1 in \cite{2017.FeRh.calorics}).
The martensitic transition is barrier\-less  (Fig.~\ref{figNEB}), with a gain of $8$~meV/atom relative to the ideal AFM(111) B2; hence, FeRh should transform to a martensite below $90\pm10\,$K.   The phonons associated with this groundstate are stable (Fig.~\ref{figPhonOrt}).  We also confirm that several structures are closely competing,  including a proposed  AFM(100) highly-distorted ($c/a$\,=\,$1.23$)  body-centered tetragonal (BCT) structure \cite{2016.PRB.94.180407}.
Our results establish that this BCT structure is not the global groundstate, but a higher-energy tetragonally constrained solution (Table~\ref{t2E}), which might be stabilized by strain.
Without constraints, this system (with an unstable M-point phonon) distorts, with accompanying atomic shuffles (Fig.~\ref{fig1} and Table~\ref{t1ort}) that stabilize the orthorhombic structure (Table~\ref{t2E}).
Albeit, due to larger entropy (Fig.~3 in \cite{2017.FeRh.calorics}), the more symmetric B2 has lower Gibbs free energy at room temperature, in agreement with the  observed austenitic AFM(111) B2 phase  \cite{PhysRev131p183y1963,Swartzendruber1984,Intermetallics15n9p1237y2007}.

{\par}Interestingly, the austenitic phases of NiTi and FeRh have the same nominal B2 structure (CsCl, $Pm\bar{3}m$ space group), and both exhibit a large caloric effect \cite{JAP37n3p1257y1966,PSSB20n1pK25y1967,Nikitin1990p363,Annaorazov1992,Annaorazov1996,JPhysD41n19p192004y2008,PRB89p214105y2014,NComms11614y2016,PRL109p255901y2012,ActaMat106p15y2016}.
Moreover, AFM B2 FeRh (below 353~K) and B2 NiTi (above 313~K) both have unstable phonon modes.
Nevertheless,
the premartensitic instability in FeRh is a surprise after all the years of experimental (e.g.,  \cite{JAP33n3p1343y1962,PhysRev131p183y1963,JETP19n6p1348y1964,PhysRev134pA1547y1964,PRB50p4196y1994,PRB81p104415y2010,PRL109p255901y2012,PRB92p184408y2015,jjimm80n3p186y2016,ActaMat106p15y2016,SciRep6p22383y2016})
and theoretical (e.g., \cite{JETP36n1p105y1973,SovPhysUsp11n5p727y1969,Hasegawa1987p175,PRB46p2864y1992,PRB46p14198y1992,Pugacheva1994p731,PRB83p174408y2011,PRB89p054427y2014,PRB91p014435y2015,PRB91p224421y2015,PRB92p094402y2015,PRB93p024423y2016,PRB94p014109y2016,PRB94p174435y2016,2016.PRB.94.180407})
studies.

{\par}  FeRh has several intriguing properties \cite{2017.FeRh.calorics,PRB94p014109y2016,PRB94p174435y2016,2016.PRB.94.180407,JAP33n3p1343y1962,JAP37n3p1257y1966,PSSB20n1pK25y1967,Nikitin1990p363,Annaorazov1992,Annaorazov1996,JPhysD41n19p192004y2008,PRL109p255901y2012,PRB89p214105y2014,NComms11614y2016,ActaMat106p15y2016,Pugacheva1994p731,PRB50p4196y1994,JETP19n6p1348y1964,PhysRev134pA1547y1964,PRB81p104415y2010,PRL109p255901y2012,PRB92p184408y2015,jjimm80n3p186y2016,ActaMat106p15y2016,SciRep6p22383y2016,JETP36n1p105y1973,SovPhysUsp11n5p727y1969,Hasegawa1987p175,PRB46p2864y1992,PRB46p14198y1992,PRB83p174408y2011,PRB89p054427y2014,PRB91p014435y2015,PRB91p224421y2015,PRB92p094402y2015,PRB93p024423y2016,PRB94p014109y2016,PRB94p174435y2016,2016.PRB.94.180407,SovPhysUsp11n5p727y1969,JAP37n3p1257y1966,JETP36n1p105y1973,JPhysC3n1SpS46y1970,PRB46p2864y1992,JAP90n12p6251y2001,PRL93p197403y2004,IEEEtransMag44n11p2875y2008,PRB81p104415y2010,PRB82p184418y2010,PRB83p174408y2011,APL104n23p232407y2014,PRB93p024423y2016,PRB93p104416y2016,PRB92p094402y2015,PRB93p064412y2016,EPL116n2p27006y2016,AIPadvances6n1p015211y2016,SciRep6p22383y2016,2017.FeRh.calorics,2016.PRB.94.180407}.
The AFM--FM metamagnetic transition is accompanied by a change of Rh moments from 0 to $\sim\!1 \, \mu_B$ (FM) \cite{2017.FeRh.calorics,2016.PRB.94.180407},
with large reversible magneto-, baro-, and elasto-caloric effects \cite{Nikitin1990p363,Annaorazov1992,Annaorazov1996,JPhysD41n19p192004y2008,PRL109p255901y2012,PRB89p214105y2014,NComms11614y2016,ActaMat106p15y2016},
anomalous structural behavior \cite{PRB92p184408y2015,PRB94p014109y2016,PRB94p174435y2016},
a  giant volume magnetostriction \cite{PRB50p4196y1994},
and a giant magnetoresistance \cite{PRB46p14198y1992,NMat13n4p367y2014}.
This transition temperature is highly sensitive to composition \cite{Swartzendruber1984,Intermetallics15n9p1237y2007,PRB89p054427y2014} and external fields \cite{Vinokurova1981,JAP113n12p123909y2013,jjimm80n3p186y2016,JPhysD49n20p205003y2016,NMat13p345y2014}.
Cooling from a melt, FeRh solidifies at $1600^\circ$C \cite{Intermetallics15n9p1237y2007}, chemically orders into B2 at $1350^\circ$C, 
magnetically orders into FM at $440^\circ$C  and AFM below $80^\circ$C, 
and transforms into a martensite at a cryogenic temperature.
Elsewhere~\cite{2017.FeRh.calorics}, we address the quantitative prediction of thermodynamic and caloric quantities associated with the AFM-FM transition.
Here we focus on determining the FeRh groundstate and assessing its structure.

\section*{\label{GS}The Groundstate }
{\par} To determine the groundstate of FeRh, along with competing magnetic structures, we constructed a set of super\-cells with the type-2 AFM ordering of the atomic magnetic moments (Fig.~1 in \cite{2017.FeRh.calorics}).
Using DFT-based \emph{ab initio} molecular dynamics, we equilibrated each super\-cell between $353$--$1000\,$K,  cooled it to $0\,$K, and performed a conjugate-gradient-based relaxation, keeping the AFM(111) spin order.
We found a stable final structure with an orthorhombic primitive cell  ($Pmmn$ space group \#59), see Fig.~\ref{fig1} and Table~\ref{t1ort}, which is $8$\,meV/atom  ($15$\,meV/FeRh) lower than the ideal  B2 (Table~\ref{t2E}).
Considering various super\-cells, we obtained convergence to the same structure (sometimes with planar defects, consistent with formation of a martensite).


{\par} The direction of atomic shuffles in this orthorhombic groundstate is consistent with an unstable M-point $(\frac{1}{2}\frac{1}{2}0)$ phonon mode, found for the AFM(111) B2 \cite{2017.FeRh.calorics,PRB94p014109y2016,2016.PRB.94.180407}.
From Table~\ref{t2E} it is clear that the AFM(100) BCT state with a large $c/a$ distortion is not the global groundstate, in contrast to the claim in \cite{2016.PRB.94.180407}.
However, if the cell is constrained such that $b\,$=$\,a$ and no atomic shuffles are permitted, then the BCT AFM(100) state is slightly (1--2$\,$meV) below  the AFM(111) orthorhombic structure.

{\par}The $D_{2h}$ $Pmmn$ symmetry (space group \#59) of electronic and atomic order accounts for different Fe-moment orientations  in the AFM structure (Fig.~\ref{fig1} and Table~\ref{t1ort}).  If only atomic ordering is considered, then the space group is $C_{2v}$ $Pmma$ (\#51).

\begin{table}[b!]
\begin{tabular}{rlcll}
\hline
$x$ &   &$y_{ort}$           & $z$ & atom \\
\hline
0    &    & $0 -d_{Fe}$      &    0  & Fe $\uparrow$ \\
0.5 &    & $0.5+d_{Fe}$   &   0.5 & Fe $\uparrow$ \\
0    &    & $0 -d_{Fe}$      &    0.5  & Fe $\downarrow$ \\
0.5 &    & $0.5+d_{Fe}$   &    0 & Fe $\downarrow$ \\
0    &    & $0.5-d_{Rh}$   &    0.25 & Rh \\
0    &    & $0.5-d_{Rh}$   &    0.75 & Rh \\
0.5 &    & $0+d_{Rh}$     &    0.25 & Rh \\
0.5 &    & $0+d_{Rh}$     &    0.75 & Rh \\
\hline
\end{tabular}
\caption{\label{t1ort} Direct (fractional) lattice coordinates of  atoms in the AFM orthorhombic ($Pmmn$)  structure  with atomic shuffles $d_{Fe}$=0.0612 and $d_{Rh}$= 0.0527
and with lattice constants $a_{ort} \! =\! 4.257 \, \mbox{\AA}$, $b_{ort} \! =\! 4.434 \, \mbox{\AA}$, and $c_{ort} \! =\! 5.584 \, \mbox{\AA}$  along $[110]$, $[\bar{1}10]$, and $[002]$ cubic directions, respectively (Fig.~\ref{fig1}).
In Cartesian coordinates $(x a_{ort}; y b_{ort}; z c_{ort})$, the shuffles are $d_{Fe} \cdot b_{ort} = 0.27\,${\AA} and $d_{Rh} \cdot b_{ort} = 0.23\,${\AA}.
AFM B2 has no shuffles, with $(a,b,c)$=$(a\sqrt{2}; a\sqrt{2}; 2a)$, where $a =2.993\,${\AA}. Arrows show relative orientation of atomic magnetic moments.
}
\end{table}

\begin{table}[b]
\begin{tabular}{lcrll}
\hline
structure & spin & $\Delta E$ && $c/a_{tet}$ \\
\hline
B2   & NM   & $1071$  && 1.0 \\
B2   & FM	 &  $60$    && 1.0 \\
BCT & (111)  & $4$	 && 1.216 \\
BCT & (100) & $1$       && 1.245 \\
B2   & (111) &  $0$	 && 1.0 \\
ort   & (111)  & $-15$   &&	0.909 \\
\hline
\end{tabular}
\caption{\label{t2E}
Energy differences $\Delta E = E - E_{B2}^{(111)}$ (meV/FeRh) and distortions $c/a_{tet} \equiv ({c^3/V})^{1/2}$ of the  cubic (B2), body-centered tetragonal (BCT), and orthorhombic (ort) structures with non-magnetic (NM), ferromagnetic (FM), or AFM spin order with a spin wave along the specified direction.}
\end{table}

{\par} The orthorhombic structure in Fig.~\ref{fig1} can be viewed as a monoclinic ($P2/m$) with a twice larger unit cell (Fig.~5 in \cite{PRB94p174435y2016}).
Figure~\ref{fig1}(b) shows the primitive orthorhombic and non-primitive monoclinic unit cells.
With $a_m\,$=$\,b_m\,$$\ne$$\,c_m$, the relation between them is clear:  $\vec{a}_m =  \vec{a}_{ort}   +    \vec{b}_{ort}$,  $\vec{b}_m =  - \vec{a}_{ort}   +   \vec{b}_{ort}$, and $\vec{c}_m =\vec{c}_{ort}$,
where  $\gamma = 87.7^\circ \! = 90^\circ \! - \!  \Delta \gamma$ is the angle between $\vec{a}_m$ and $\vec{b}_m$.
This structure differs from a tetragonal one by $\Delta \gamma = 2.3^\circ$.  Importantly, the atomic shuffles  ($0.27~\mbox{\AA}$ for Fe and $0.23~\mbox{\AA}$ for Rh, see Table~\ref{t1ort})   destroy tetragonal symmetry. Shuffles along the [110] direction (Fig.~\ref{fig1}) can be compared with the superposition of two degenerate unstable phonon modes at $M$, shown in Fig.~3(d) in \cite{2016.PRB.94.180407}.

\section*{\label{MartensiticNEB} Martensitic Transformation}

\begin{figure}[t]
\includegraphics[width=70mm]{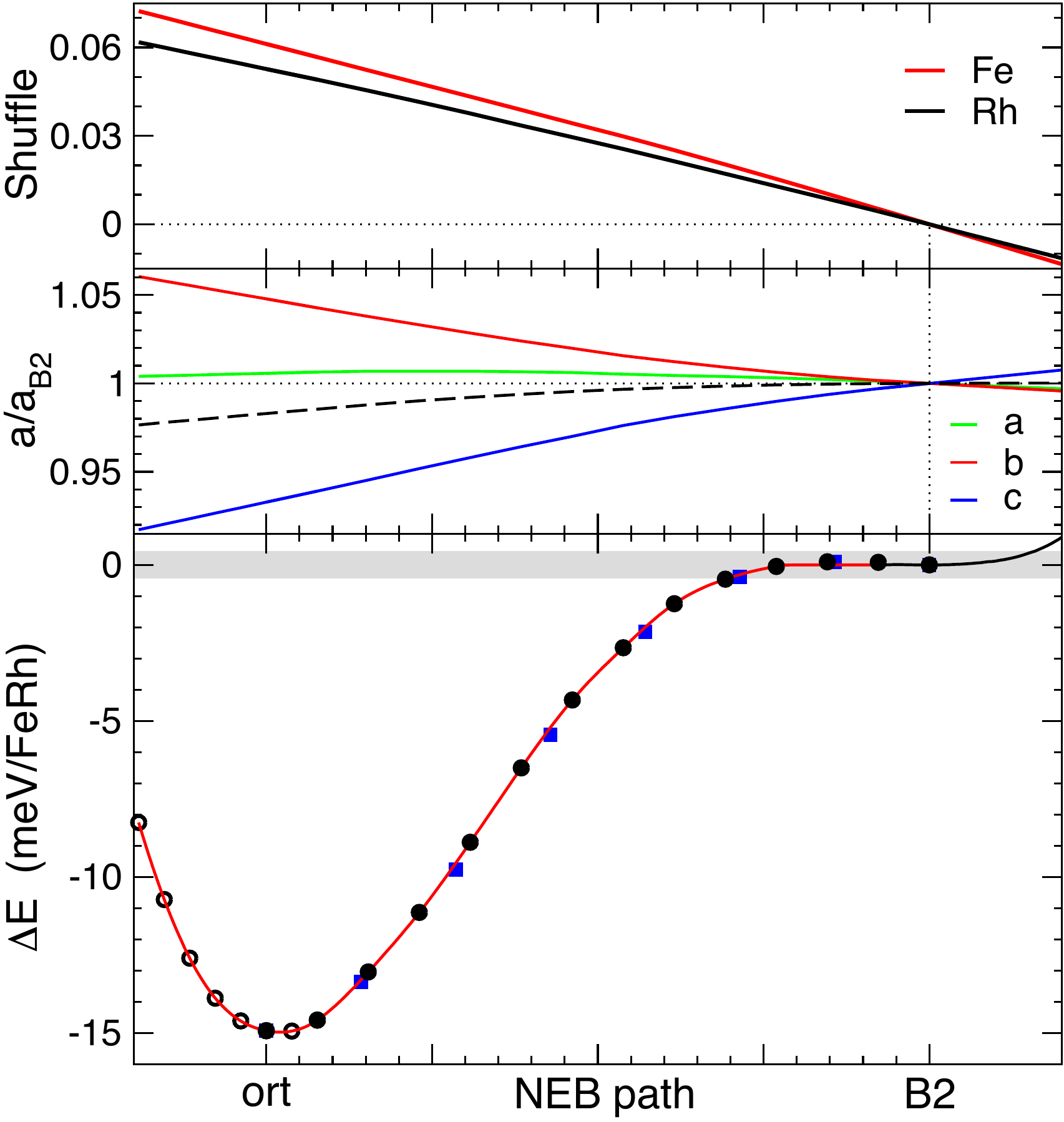}  
\caption{\label{figNEB} (Color online). MEP for B2-to-ortho given by SSNEB:
atomic shuffles $d = y_{ort} - y_{B2}$ (see Table~\ref{t1ort}), distortions of the lattice constants (solid lines) and volume (dashed line), and
energies $\Delta E$ (relative to B2). No barrier is evident.  For the 8-atom unit cell (Table~\ref{t1ort}), $\Delta E$ from 12-image SSNEB (filled circles)   is accompanied by a linear continuation of MEP from the terminal images and their neighbors (open circles near ortho and black line near B2; red line is the cubic spline), and compared to 6-image SSNEB (blue squares). 
}
\end{figure}

{\par} Our SSNEB results (Fig.~\ref{figNEB}) directly confirm that this AFM(111) orthorhombic structure is stable  and $8 \,$meV/atom lower than B2 austenite.  The orthorhombic phase is  anisotropic and can form a martensite.
The MEP in Fig.~\ref{figNEB} is characterized by the coupling of atomic shuffles and lattice deformations. Moreover, there is no enthalpy barrier. 
From the SSNEB calculations in a small 8-atom cell 
and a larger 16-atom cell, 
we found that the MEP can be described by only 5 degrees of freedom: 2 atomic shuffles and 3 lattice constants (Fig.~\ref{figNEB}).

{\par} From the energy gain $\Delta E=(E^{B2}-E^{ ort})$, we predict the martensitic temperature $T_m$ at $90\pm10\,$K, estimated from $T_m = \Delta E/ k_B$, see eq.~3.7 in \cite{ZarkevichPhD2003}.
This estimate is approximate; it does not take into account zero-point vibrations
(significant for hydrogen and light elements, but small for heavy $_{\,\,\,45}^{103}$Rh and $_{26}^{56}$Fe)
and excitations of the internal degrees of freedom, such as phonons or magnons.
Due to sensitivity to strain, $T_m$ could  be suppressed  by the martensitic stress \cite{PRB91p174104}.
Indeed, constraints on any degrees of freedom during the martensitic transformation (Fig.~\ref{figNEB})
lead to an under-relaxed structure with a higher energy $E$ and a reduced energy gain $(E^{B2}-E)$.
For example, atomic shuffles alone give the B2 M-point phonon instability at $\sim$20~K.
Hence, stress or constraints on the lattice constants (present in experimental samples that are not single crystals)
could suppress  $T_m$ significantly.  
Without constraints, the relative energies of the competing phases are given in Table~\ref{t2E},
and the BCT solution (A$^\prime$-AFM in \cite{2016.PRB.94.180407}) is not the global groundstate.


\section*{\label{a4}Phonons}
{\par} Although the AFM-B2 austenite is stabilized by entropy at room temperature, it should have a pre\-martensitic instability, similar to that in NiTi austenite \cite{PRB90p060102y2014,PRL113p265701y2014}.
This  anomalous structural behavior of austenite is indeed indicated by its phonons: Fig.~2 in \cite{2017.FeRh.calorics} shows a sensitivity to atomic displacements and their magnitude.  This instability is small compared to the pre\-martensitic instability in NiTi, see Fig.~2a in \cite{PRB90p060102y2014}.

\begin{figure}[t]
\includegraphics[width=75mm]{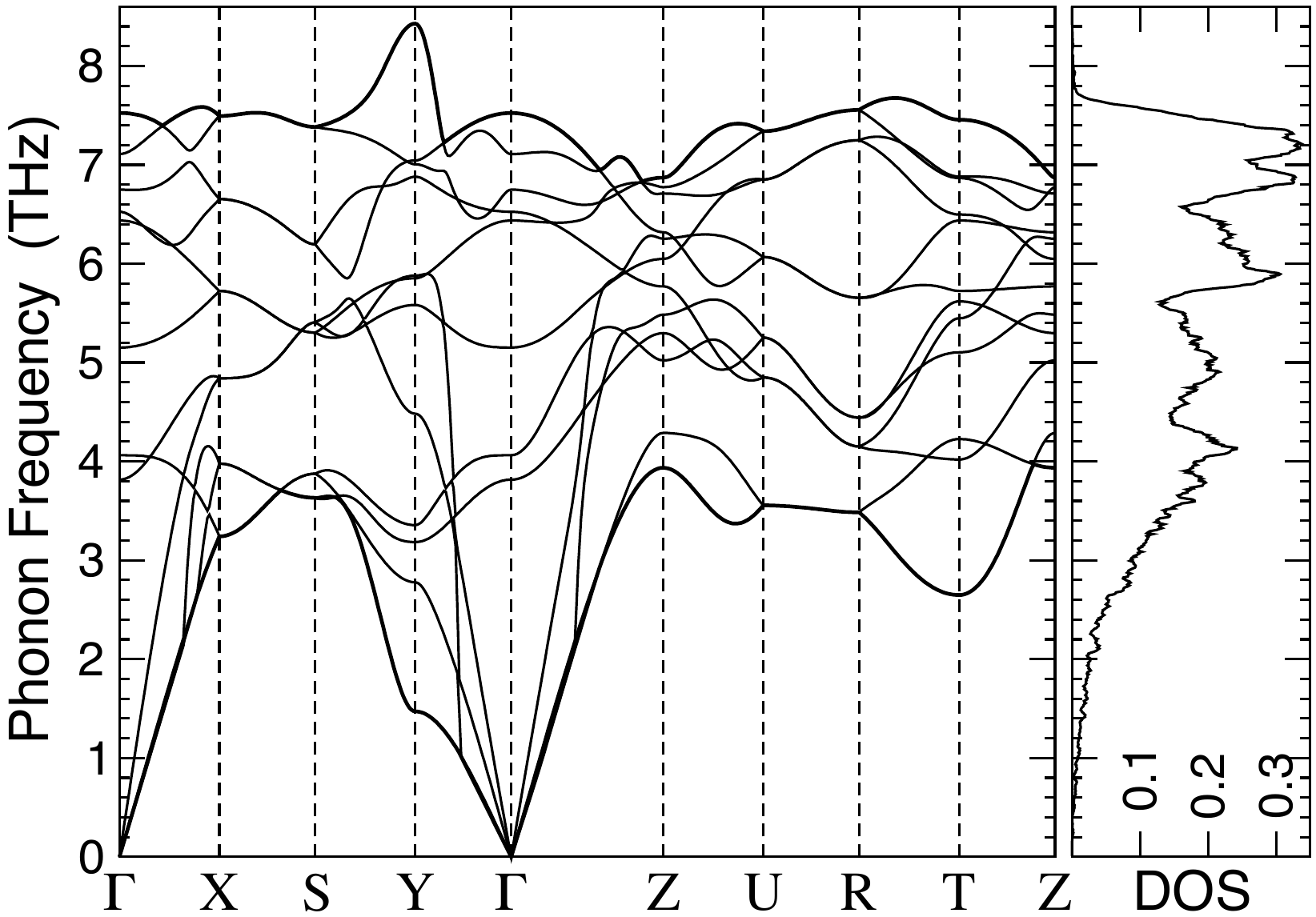}  
\caption{\label{figPhonOrt} Phonon frequencies and DOS for the ortho\-rhombic AFM FeRh groundstate (Fig.~\ref{fig1}). }
\end{figure}

{\par} Using the small-displacement method \cite{Phon} at zero pressure, we find an unstable  phonon mode at $M$ $(\frac{1}{2} \frac{1}{2} 0)$ in the B2 AFM phase,  but not in the FM phase \cite{2017.FeRh.calorics};  
we agree with experiment \cite{PRB92p184408y2015} and recent calculations \cite{2016.PRB.94.180407,PRB94p174435y2016,PRB94p014109y2016}.
However, the predicted AFM ortho\-rhombic groundstate is stable, and has stable phonons  (Fig.~\ref{figPhonOrt}).
The groundstate structure has a lattice translation vector (Fig.~\ref{fig1}) and atomic shuffles (Table~\ref{t1ort}) along the cubic [110] direction, consistent with the $M$-point distortion of B2.

\section*{\label{Summary}Summary}
{\par} We confirmed that the  cubic B2 FeRh structure in AFM(111) state is unstable with respect to infinitesimal atomic displacements corresponding to $M$-point phonons.
We predicted an AFM(111) orthorhombic groundstate structure (Tables~\ref{t1ort}~and~\ref{t2E}), consistent with an M-point distortion of B2, that is  8~meV/atom below B2; thus, a martensitic transformation is expected below $90\pm10$\,K.
From the solid-state nudged elastic band calculations, we showed that the minimum-enthalpy pathway from B2 to the orthorhombic groundstate has no barrier.
However, as is well established, martensitic stresses can suppress such transitions to lower temperatures.
This symmetry-breaking transformation from cubic austenite to anisotropic martensite involves both finite lattice distortions and atomic shuffles. However,  the shuffled local minima of the  potential energy in the martensite are very shallow, so the thermal atomic motion can be harmonic only below $23\pm10$\,K.
We look forward to experimental verification of the predicted FeRh martensitic transformation and its low-$T$ structure.

{\par}We also confirmed the existence of other competing low-energy structures, such as the AFM(100) BCT, suggested as a groundstate  \cite{2016.PRB.94.180407};  this structure can be stabilized by strain (i.e., tetragonal constraint), which might occur in thin-film grown samples, for example.
Nonetheless, we find that its energy is above that of the unrestricted AFM(111) orthorhombic structure.

{\par}
The thermodynamics associated with the phase transitions of the low-$T$ martensitic
and ambient-$T$ 
meta\-magnetic (predicted at $346\,$K and observed at $353\,$K)
have been explored in details elsewhere \cite{2017.FeRh.calorics},
where we
use estimators to evaluate transition temperatures, caloric effects,  specific heat,  entropy, and thermal expansion  with very good accuracy.

\acknowledgments
Applications to caloric materials discovery at Ames Laboratory is supported by the U.S. Department of Energy (DOE),  Advanced Manufacturing Office of the Office of Energy Efficiency and Renewable Energy through CaloriCool\textsuperscript{TM} -- the Caloric Materials Consortium established as a part of the U.S. DOE Energy Materials Network.  In part, predictive methods development was funded by the U.S. DOE, Office of Science, Basic Energy Sciences, Materials Science and Engineering Division. Ames Laboratory is operated for the U.S. DOE by Iowa State University under contract DE-AC02-07CH11358.

\section*{\label{ComputationalDetails}Computational Details}
DFT calculations were performed via a modified VASP code  \cite{VASP1,VASP2} with built-in double climbing image (C2NEB) algorithm \cite{C2NEB,C2NEBsoft}.
The solid-state nudged elastic band method without climbing (preserved in C2NEB code \cite{C2NEBsoft}) was used to find the barrierless transformation (Fig.~\ref{figNEB}).
We used the projector augmented waves (PAW) basis \cite{PAW,PAW2} and PBE exchange-correlation functional \cite{PBE} with Vosko-Wilk-Nusair spin-polarization \cite{VOSKOWN}.
The PAW-PBE potentials with 8 ($d^7 s^1$) and 9 ($d^8s^1$) valence electrons (Ar and Kr cores) were used for Fe and Rh, respectively.
The Brillouin zone integration was performed on a dense Monkhorst-Pack mesh \cite{MonkhorstPack1976} with $\ge 50$ $k$-points per {\AA}$^{-1}$, including $\Gamma$.
An additional, third support grid was used for the evaluation of the augmentation charges.
The plane-wave energy cut-off was set to 334.9 eV; the variation of the energy differences at higher cut-offs (up to 4000 eV) did not exceed 0.5 meV per formula unit.

{\par} The \textit{ab initio} molecular dynamics (MD) followed by full relaxation (using the conjugate-gradient algorithm) allowed to find the lowest-energy structure.
MD with 2~fs time steps was performed in several periodic boxes of various sizes, including cubic 16-atom $2 \times 2 \times 2$ and 128-atom $4 \times 4 \times 4$.
Symmetry was not preserved during MD.
The initial B2 structure with type-2 AFM spin order was relaxed, then equilibrated using MD (at fixed $T$ between 353 and $1000$\,K, fixed volume and the number of atoms),
after that quenched to $0\,$K,
and fully relaxed again.
The equilibration times varied from 4000 to 24000~fs. 
In our searches, multiple replicas of each box were quenched and relaxed during continued equilibration.
For each box size, we tried to vary both equilibration time and cooling rate before the final relaxation.
For example, a 16-atom system was equilibrated at $400\,$K and quenched at the cooling rate of 0.1 K/fs (hence, $T$ linearly changed from 400~K to $0\,$K during 4000~fs).
The final full relaxation was done in either 1 or 3 steps, using one of the following three algorithms:
(1) simultaneously relax both lattice constants and atomic positions;
(2) relax atoms at fixed volume, than relax volume with fixed atomic positions, and finally relax all degrees of freedom;
(3) relax lattice constants, than atoms, than all degrees of freedom.
In the algorithms (2) and (3), the last step is identical to (1).
We compared energies (per atom) of the final structures in multiple simulation boxes of various dimensions and after various equilibration, cooling, and relaxation procedures.
The lowest-energy structure is in Fig.~\ref{fig1} and Table~\ref{t1ort}.

{\par}
If there are multiple local minima on the potential energy surface, then molecular dynamics at a low $T$ (as well as a structural relaxation) can be trapped in one of those.
Also, if the periodic boundary conditions are incommensurate with the periodicity of the groundstate, then the simulated annealing is prevented from converging to the lowest-energy structure.
That is why we considered various box sizes and quenched multiple replicas of the equilibrated structures at various cooling rates.
We found that a number of simulations converged to the same periodic structure (Fig.~\ref{fig1}), which had the lowest energy per atom.
The other final structures had higher energies; most of them differed from the groundstate by twins and defects.

{\par}
Phonons were calculated using the small-displacement method, implemented in the \emph{Phon} code \cite{Phon}.
In a cubic $4 \!\times\! 4 \!\times\! 4$ supercell containing 64 FeRh formula units (128 atoms),
we used 6 displacements of atoms by 0.04 {\AA};
these displacements are small compared to the shuffles of Fe ($0.27\,${\AA})
and Rh ($0.23\,${\AA}),
see Table~\ref{t1ort}.

\bibliography{FeRh}

\end{document}